\documentclass[prl,unsortedaddress,superscriptaddress,showpacs,twocolumn,floatfix]{revtex4}

\usepackage{pstcol}
\usepackage{graphicx}

\begin{document}

\title{Measurement of the Muon Decay Parameter $\delta$}

\affiliation{University of Alberta, Edmonton, AB, T6G 2J1, Canada}
\affiliation{University of British Columbia, Vancouver, BC, V6T 1Z1, Canada}
\affiliation{Kurchatov Institute, Moscow, 123182, Russia}
\affiliation{University of Montreal, Montreal, QC, H3C 3J7, Canada}
\affiliation{University of Regina, Regina, SK, S4S 0A2, Canada}
\affiliation{Texas A\&M University, College Station, TX 77843, U.S.A.}
\affiliation{TRIUMF, Vancouver, BC, V6T 2A3, Canada}
\affiliation{Valparaiso University, Valparaiso, IN 46383, U.S.A.}

\author{A.~Gaponenko}
\affiliation{University of Alberta, Edmonton, AB, T6G 2J1, Canada}

\author{R.~Bayes}
\altaffiliation[Affiliated with: ]{Univ.~of Victoria,
Victoria, BC.}
\affiliation{TRIUMF, Vancouver, BC, V6T 2A3, Canada}

\author{Yu.I.~Davydov}
\altaffiliation[Affiliated with: ]{Kurchatov Institute,
Moscow, Russia.}
\affiliation{TRIUMF, Vancouver, BC, V6T 2A3, Canada}

\author{P.~Depommier}
\affiliation{University of Montreal, Montreal, QC, H3C 3J7, Canada}

\author{J.~Doornbos}
\affiliation{TRIUMF, Vancouver, BC, V6T 2A3, Canada}

\author{W.~Faszer}
\affiliation{TRIUMF, Vancouver, BC, V6T 2A3, Canada}

\author{M.C.~Fujiwara}
\affiliation{TRIUMF, Vancouver, BC, V6T 2A3, Canada}

\author{C.A.~Gagliardi}
\affiliation{Texas A\&M University, College Station, TX 77843, U.S.A.}

\author{D.R.~Gill}
\affiliation{TRIUMF, Vancouver, BC, V6T 2A3, Canada}

\author{P.~Green}
\affiliation{University of Alberta, Edmonton, AB, T6G 2J1, Canada}

\author{P.~Gumplinger}
\affiliation{TRIUMF, Vancouver, BC, V6T 2A3, Canada}

\author{M.D.~Hasinoff}
\affiliation{University of British Columbia, Vancouver, BC, V6T 1Z1, Canada}

\author{R.S.~Henderson}
\affiliation{TRIUMF, Vancouver, BC, V6T 2A3, Canada}

\author{J.~Hu}
\affiliation{TRIUMF, Vancouver, BC, V6T 2A3, Canada}

\author{B.~Jamieson}
\affiliation{University of British Columbia, Vancouver, BC, V6T 1Z1, Canada}

\author{P.~Kitching}
\affiliation{University of Alberta, Edmonton, AB, T6G 2J1, Canada}

\author{D.D.~Koetke}
\affiliation{Valparaiso University, Valparaiso, IN 46383, U.S.A.}

\author{A.A.~Krushinsky}
\affiliation{Kurchatov Institute, Moscow, 123182, Russia}

\author{Yu.Yu.~Lachin}
\affiliation{Kurchatov Institute, Moscow, 123182, Russia}

\author{J.A.~Macdonald}
\altaffiliation[Deceased.]{} 
\affiliation{TRIUMF, Vancouver, BC, V6T 2A3, Canada}

\author{R.P.~MacDonald}
\affiliation{University of Alberta, Edmonton, AB, T6G 2J1, Canada}

\author{G.M.~Marshall}
\affiliation{TRIUMF, Vancouver, BC, V6T 2A3, Canada}

\author{E.L.~Mathie}
\affiliation{University of Regina, Regina, SK, S4S 0A2, Canada}

\author{L.V.~Miasoedov}
\affiliation{Kurchatov Institute, Moscow, 123182, Russia}

\author{R.E.~Mischke}
\affiliation{TRIUMF, Vancouver, BC, V6T 2A3, Canada}

\author{J.R.~Musser}
\affiliation{Texas A\&M University, College Station, TX 77843, U.S.A.}

\author{P.M.~Nord}
\affiliation{Valparaiso University, Valparaiso, IN 46383, U.S.A.}

\author{M.~Nozar}
\affiliation{TRIUMF, Vancouver, BC, V6T 2A3, Canada}

\author{K.~Olchanski}
\affiliation{TRIUMF, Vancouver, BC, V6T 2A3, Canada}

\author{A.~Olin}
\altaffiliation[Affiliated with: ]{Univ.~of Victoria,
Victoria, BC.}
\affiliation{TRIUMF, Vancouver, BC, V6T 2A3, Canada}

\author{R.~Openshaw}
\affiliation{TRIUMF, Vancouver, BC, V6T 2A3, Canada}

\author{T.A.~Porcelli}
\altaffiliation[Present address: ]{Univ.~of Manitoba,
Winnipeg, MB.}
\affiliation{TRIUMF, Vancouver, BC, V6T 2A3, Canada}

\author{J.-M.~Poutissou}
\affiliation{TRIUMF, Vancouver, BC, V6T 2A3, Canada}

\author{R.~Poutissou}
\affiliation{TRIUMF, Vancouver, BC, V6T 2A3, Canada}

\author{M.A.~Quraan}
\affiliation{University of Alberta, Edmonton, AB, T6G 2J1, Canada}

\author{N.L.~Rodning}
\altaffiliation[Deceased.]{} 
\affiliation{University of Alberta, Edmonton, AB, T6G 2J1, Canada}

\author{V.~Selivanov}
\affiliation{Kurchatov Institute, Moscow, 123182, Russia}

\author{G.~Sheffer}
\affiliation{TRIUMF, Vancouver, BC, V6T 2A3, Canada}

\author{B.~Shin}
\altaffiliation[Affiliated with: ]{Univ.~of Saskatchewan,
Saskatoon, SK.}
\affiliation{TRIUMF, Vancouver, BC, V6T 2A3, Canada}

\author{F.~Sobratee}
\affiliation{University of Alberta, Edmonton, AB, T6G 2J1, Canada}

\author{T.D.S.~Stanislaus}
\affiliation{Valparaiso University, Valparaiso, IN 46383, U.S.A.}

\author{R.~Tacik}
\affiliation{University of Regina, Regina, SK, S4S 0A2, Canada}

\author{V.D.~Torokhov}
\affiliation{Kurchatov Institute, Moscow, 123182, Russia}

\author{R.E.~Tribble}
\affiliation{Texas A\&M University, College Station, TX 77843, U.S.A.}

\author{M.A.~Vasiliev}
\affiliation{Texas A\&M University, College Station, TX 77843, U.S.A.}

\author{D.H.~Wright}
\altaffiliation[Present address: ]{Stanford Linear Accelerator Center,
Stanford, CA.}
\affiliation{TRIUMF, Vancouver, BC, V6T 2A3, Canada}

\collaboration{TWIST Collaboration}
\noaffiliation

\date{April 4, 2005}

\begin{abstract}
The muon decay parameter $\delta$ has been measured by the TWIST
collaboration.  We find $\delta$ = 0.74964 $\pm$ 0.00066(stat.) $\pm$
0.00112(syst.), consistent with the Standard Model value of 3/4.  This result
implies that the product $P_{\mu}\xi$ of the muon polarization in pion decay,
$P_{\mu}$, and the muon decay parameter $\xi$ falls within the 90\% confidence
interval $0.9960 < P_{\mu}\xi \leq \xi < 1.0040$.  It also has implications
for left-right-symmetric and other extensions of the Standard Model.
\end{abstract}

\maketitle
The TWIST spectrometer \cite{Henderson} was designed to
measure a broad range of the normal muon decay spectrum,
 $\mu^+ \rightarrow e^+ \nu_e \overline{\nu}_{\mu}$,
allowing the simultaneous extraction of the spectrum shape
parameters.   Assuming the weak interaction is local and invariant under the 
Lorentz group, the effective four fermion muon decay matrix element can be
written in terms of helicity-preserving amplitudes:
\begin{equation}
M = \frac{4 G_F}{\sqrt{2}} \sum_{\gamma=S,V,T;\epsilon,\mu=R,L} 
g^{\gamma}_{\epsilon\mu}
\langle \bar{e}_{\epsilon} | \Gamma^{\gamma} | \nu \rangle
\langle \overline{\nu} | \Gamma_{\gamma} | \mu_{\mu} \rangle ,
\end{equation}
where the $g^{\gamma}_{\epsilon\mu}$ specify the scalar, vector, and tensor
couplings between $\mu$-handed muons and $\epsilon$-handed electrons
\cite{Fetscher}.  In this form, the Standard Model implies $g^V_{LL}$ = 1 and
all other coupling constants are zero.

The differential decay spectrum \cite{Michel} of the $e^{+}$ emitted in the
decay of polarized $\mu^{+}$ is provided in terms of four parameters, $\rho$,
$\delta$, $\eta$, and $\xi$, commonly referred to as the Michel parameters,
which are bilinear combinations of the coupling constants.  In the limit where
the electron and neutrino masses as well as radiative corrections are
neglected, this spectrum is given by:
\begin{eqnarray}
    \frac{d^2 \Gamma}{x^2 dx d(\cos \theta)} & \propto &
3 (1-x)  +  \frac{2}{3} \rho (4x-3)  \nonumber \\
& + &  P_{\mu} \xi \cos \theta [ 1 - x + \frac{2}{3} \delta (4x-3) ],
\label{eq:decayrate}
\end{eqnarray}
where $\theta$ is the angle between the muon polarization and the outgoing
electron direction, $x=E_e/E_{max}$, and $P_{\mu}$ is the muon
polarization. The fourth parameter, $\eta$, appears in the isotropic term when
the electron mass is included in the analysis.  In the Standard Model, the
Michel parameters take on precise values.

The parameter $\xi$ expresses the level of parity violation in muon decay,
while $\delta$ parametrizes its momentum dependence. Recently, TWIST reported
a new measurement of $\rho$ \cite{Musser}.  In this paper we report a new
measurement of $\delta$.  The currently accepted value of $\delta$ =
$0.7486\pm0.0026\pm0.0028$ \cite{Balke} agrees with the Standard Model
expectation of 3/4.  Some Standard Model extensions require deviations from
pure $V-A$ coupling that can alter $\delta$.  Some of these models involve
right-handed interactions. The positive definite quantity,
\begin{eqnarray}
Q^{\mu}_{R} & = & \frac{1}{4} |g^S_{LR}|^2 + \frac{1}{4} |g^S_{RR}|^2
+ |g^V_{LR}|^2 + |g^V_{RR}|^2 + 3 |g^T_{LR}|^2 \nonumber \\
 & = & \frac{1}{2} [ 1 + \frac{1}{3} \xi - \frac{16}{9} \xi\delta ] ,
\label{eq:RHcurrents}
\end{eqnarray}
can serve to set a model independent limit on any muon right-handed couplings
\cite{Fetscher,PDG}. A recent review of muon decay is presented in
\cite{Kuno}.

 Highly polarized surface muons \cite{Pifer} are
delivered to the TWIST spectrometer \cite{Henderson} from the M13
channel at TRIUMF.  The spectrometer consists of a detector made up of
56 very thin high precision chamber planes, all mounted
perpendicularly to a solenoidal 2 T magnetic field. The muons enter
this array of chambers through a $195\ \mu$m scintillator that acts as
the event trigger.  More than 80\% of the muons come to rest in the
central stopping target, which also acts as the cathode plane for the
Multi Wire Proportional Chambers (MWPC) on either side.
The decay positrons spiral through the chambers
producing hits on the wires that are recorded by time to digital converters.
These helical tracks are later analyzed to determine
precisely the positron energy and angle.  The observed momentum
resolution is 100~keV/c \cite{Musser}.  The $\cos\theta$ resolution
derived from Monte Carlo (MC) is about 0.005.  The reconstruction is
similar to \cite{Musser}, except for some details discussed
below.

TWIST determines the Michel parameters by fitting two-dimensional histograms
of reconstructed experimental decay positron momenta and angles with
histograms of reconstructed Monte Carlo data. This approach has several
advantages.  First, spectrum distortions introduced by the event
reconstruction largely cancel because MC and experimental data are analyzed
identically.  Second, because the MC simulates the detector response well, no
explicit corrections of the result are required.  Third, a blind analysis of
the result is straightforward.  It is implemented by utilizing hidden Michel
parameters $\rho_H, \delta_H$, and $\xi_H$ to generate the theoretical decays.
The decay rate can be written as
\[
\left. \frac{d^2 \Gamma}{dx d(\cos \theta)}\right|_{\rho_H, \delta_H, \xi_H}
+ \sum_{\lambda=\rho, \xi, \xi\delta} \frac{\partial}{\partial\lambda} \left[
  \frac{d^2 \Gamma} {dx d(\cos \theta)}\right] \Delta \lambda 
\]
since the decay spectrum is linear in the shape parameters.  The sum of MC
spectra is fit to the data spectrum by adjusting the $\Delta \lambda$.
$\delta$ is extracted as $(\xi_H\delta_H +
\Delta(\xi\delta))/(\xi_H+\Delta\xi)$.  
Since the hidden parameters were allowed to only deviate from their
Standard model values by no more than 0.03 it was sufficient for the
extraction of systematic uncertainties to assume that they had their
Standard model values during the blind stage of the analysis.
The MC spectra were generated including full $O(\alpha)$ radiative corrections
with exact electron mass dependence, leading and next-to-leading logarithmic
terms of $O(\alpha^2)$, leading logarithmic terms of $O(\alpha^3)$,
corrections for soft pairs, virtual pairs, and an ad-hoc exponentiation
\cite{Arbuzov}. Because TWIST at the present stage could
not provide an improved
measurement of eta, we set it, for MC spectra production, to its
current highest precision value of $-0.007$ \cite{PDG} in order
to constrain other parameters better.  The uncertainty of $0.013$
on the accepted value of $\eta$ gives a negligible uncertainty on the final
value of $\delta$.

\begin{figure}
\center{\includegraphics[width=8.6cm]{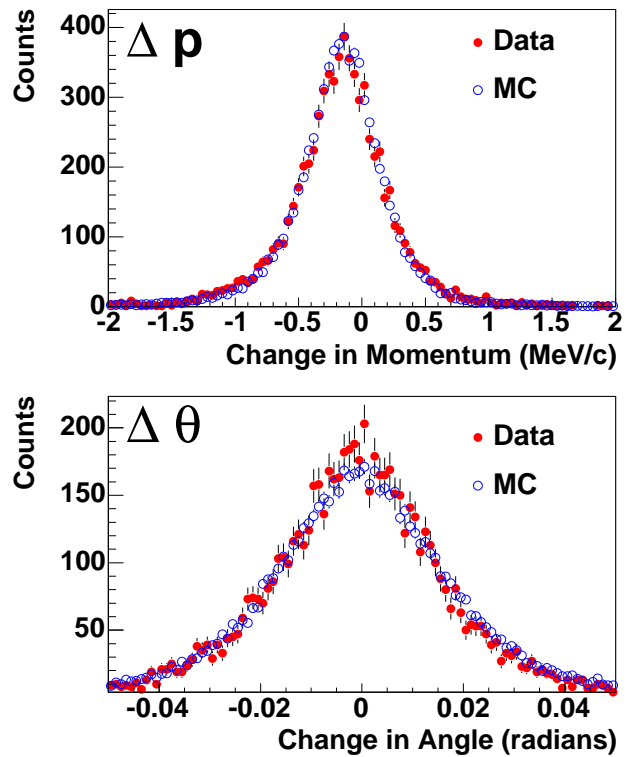}}
\caption{\label{fig:multscat} (color online) The difference between downstream
and upstream tracks, for both data and MC, resulting in: Top, the positron
momentum change in the central stopping target, Bottom, $\Delta\theta$ for a
positron that passed through the central stopping target. The MC results were
normalized to those of the data for the purposes of this figure.}
\end{figure}

The TWIST simulation model is based on GEANT 3.21 \cite{Brun} with the chamber
response based on GARFIELD \cite{garfield}. It contains virtually all the
components of the spectrometer with which a muon or a decay positron could
interact. The output exactly mimics the binary files generated
by the data acquisition system.

Factors that influence the momentum and angle determination must be well
simulated in the MC, so special runs were taken specifically to address the
accuracy of the simulation of energy loss and multiple scattering. Muons were
stopped in the extreme upstream wire chambers in both the experiment and in
the MC simulation.
The decay positrons were tracked through the upstream  half and separately
through the downstream half of the spectrometer. 
Differences in momentum and angle were
histogrammed on a track by track basis. Figure \ref{fig:multscat} presents,
for both data and MC, 
the changes in momentum and angle that occur primarily at the central stopping
target.  The widths of the peaks in this figure do not represent
the experimental
resolution for a number of reasons. First, because the same track is being
reconstructed twice with finite resolution the differences in the measured
values can be either positive or negative. Second, the particle sees
approximately twice the thickness of materials. As well, the tracking in
the upstream region is in the opposite direction for which the code is
optimized and in this region the track does not see as many planes due to
the distribution of the muon stops in the upstream planes. The MC
nonetheless reproduces the data very well.  The $\Delta p$ distribution
mean(RMS) for the data and MC are -0.17(0.41) MeV/c and -0.17(0.39) MeV/c,
respectively. The $\Delta\theta$ mean(RMS) for the data and MC are -0.95(17.0)
and -0.37(18.0) milliradians respectively.  The small differences are within
the uncertainties associated with positron interactions and target thickness.

The result for $\delta$ presented here employed a sample consisting of
$6\times10^{9}$ events recorded in Fall, 2002. This data sample is
comprised of the same 16 data sets used
for our extraction of $\rho$ \cite{Musser}. Many of these data sets
were taken under conditions chosen to
establish the sensitivity of the detector to systematic effects. Four
of the data sets, sets A and B taken at 2.00 T six weeks apart and two
other sets, one taken at 1.96 T and
one at 2.04 T, were analyzed and fit to their
corresponding MC samples to derive the value of $\delta$.
Our $\rho$ determination also utilized a cloud muon sample \cite{Musser}.
The low
polarization of that data set leads to low sensitivity and the potential for
substantially increased systematic effects in the extraction of $\delta$. 
Rather than perform a complete additional systematics study for a data set that
would contribute little weight, we chose to use the cloud muon sample only as a
consistency check on our final result for $\delta$.

There are several differences between our previous analyses for 
$\rho$ \cite{Musser} and for the $\delta$ result presented here.  After the 
first analysis was completed and the hidden
parameters were disclosed, an \textit{a priori} defined consistency
check was carried out. Muon decay parameters determined from set B were used
to generate a new Monte Carlo spectrum, which was used to perform another fit to
that set. This fit was expected to yield deviations of all parameters
consistent with zero, but it failed for $\delta$. It was determined then that
there was a flaw in the way
the polarization-dependent radiative corrections were implemented in the Monte Carlo
event generator. Since $\rho$ is essentially decoupled from the asymmetry parameters
this flaw had no impact on the value of $\rho$. After the first
analysis, we were no longer ``blind'' to the value of $\rho$.  However the flaw
did introduce a systematic uncertainty in the value of $\delta$ inferred from
the data.  This systematic uncertainty depended on the difference in absolute
polarization (not $P_{\mu}\xi$) between the actual muon decay data and the Monte Carlo
decay events.  Estimates indicated that the systematic effect was $<$\ 0.001. 
Unfortunately, a precise value was impossible to determine \textit{a priori},
as only the product $P_{\mu}\xi$ is measurable from the data, rather than
$P_{\mu}$ alone.

A new
analysis with a corrected event generator was therefore undertaken for $\delta$
through the generation, with a new set of hidden Michel parameters, of a completely
new set of Monte Carlo events. The four data sets described above were also reanalyzed
with improved alignment calibrations. Finally, the track-selection algorithm was 
improved by merging those used for the $\rho$ analysis. Thus the fits employed for the
$\delta$ extraction are completely distinct from those in \cite{Musser}.

\begin{figure}
\center{\includegraphics*[width=8.6cm]{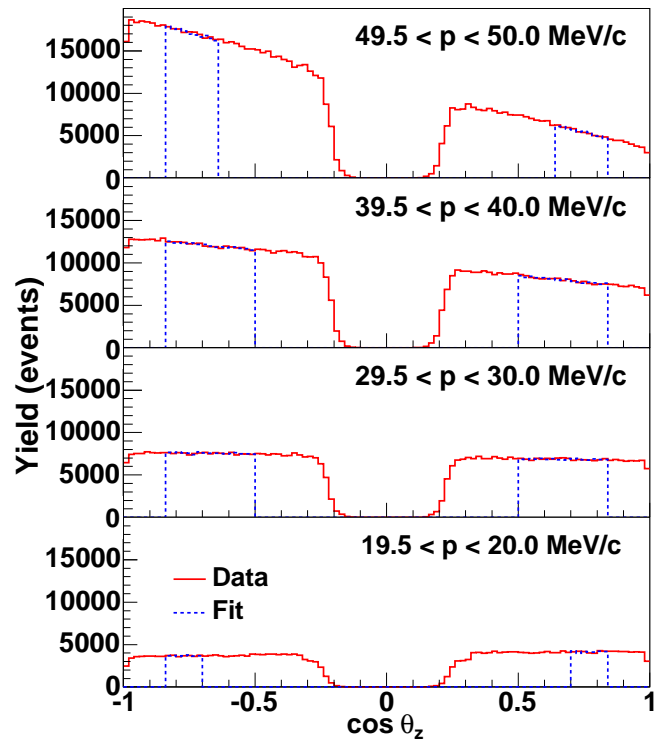}}
\caption{\label{fig:ang_dist} (color online) Decay positron angular
distributions from set B (solid curves) and the corresponding best fit
distributions within the fiducial region (dashed curves) for selected momentum
bins. $\theta_z=\pi-\theta$.}
\end{figure}

\begin{figure}
\center{\includegraphics*[width=8.6cm]{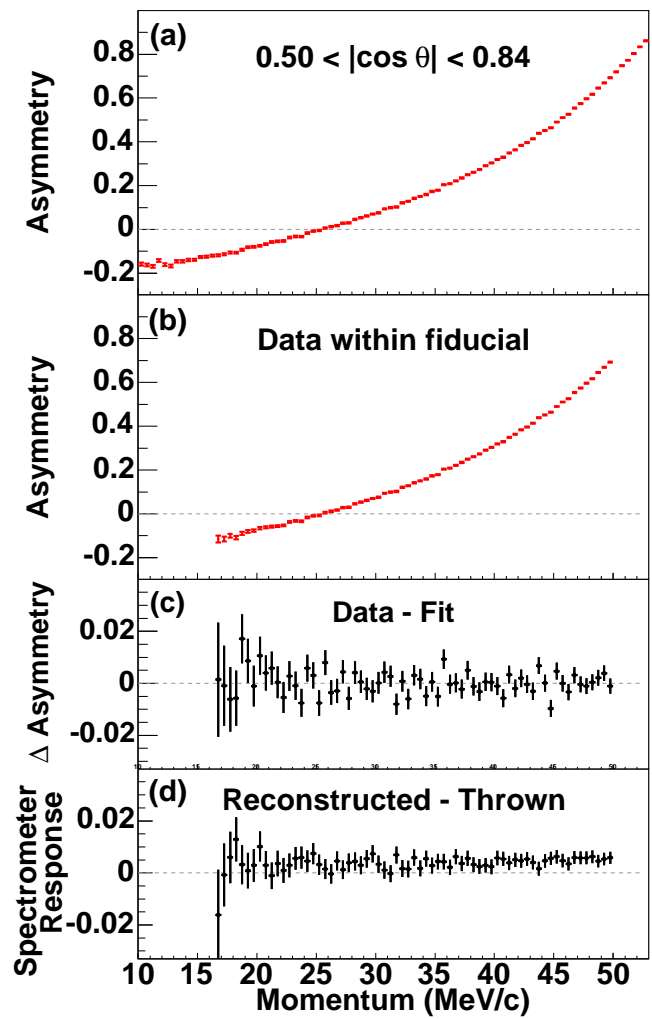}}
\caption{\label{fig:mc_data} (color online) (a) The observed muon decay
asymmetry from set B for all events within $0.50 < |\cos\theta| < 0.84$.
(b) The same quantity for those events that fall within the fiducial region.
(c) The difference between the data in panel (b) and the best fit MC spectrum.
(d) The difference between the asymmetry calculated from the reconstructed MC
events and as thrown by the MC, which illustrates the spectrometer response.}
\end{figure}

Figure \ref{fig:ang_dist} shows the decay positron angular distributions for
representative momentum bins.  Equation (\ref{eq:decayrate}) indicates the
angular distributions follow a $1 + A(p)\cos\theta$ shape, where by convention
the asymmetry, $A(p)$, is positive when positrons are emitted preferentially
along the muon polarization axis.
Figure \ref{fig:mc_data}(a) shows the observed muon decay asymmetry as a
function of momentum for set B.
The asymmetry provides a compact representation of the angular distributions.
However, extracting $\delta$ from $A(p)$ involves a significant correlation
between $\delta$ and $\rho$ \cite{Balke}.  In contrast, we extract $\delta$
from a simultaneous fit of the full experimental momentum-angle distribution
illustrated in Fig. \ref{fig:ang_dist}, as described above, which leads to a
negligible correlation between $\delta$ and $\rho$.
Fits to upstream minus downstream distributions, which are essentially
independent of $\rho$ and $\eta$, gave nearly identical results for $\delta$.

The fiducial region adopted for this analysis requires $p<50$ MeV/c,
$|p_z|>13.7$ MeV/c, $p_T<38.5$ MeV/c, and $0.50<|\cos\theta|<0.84$.  
The fiducial cuts, while intentionally chosen to be conservative,
are related to physical limitations of the TWIST
detector. The 50 MeV momentum cut rejects events that are near the region
utilized in the end point fits \cite{Musser}.  It is also important to avoid the
region very close to the end point to minimize the sensitivity of the
Michel parameter fits to the momentum resolution.
The longitudinal momentum constraint eliminates events with wavelengths
that match a 12.4 cm periodicity in the wire chamber construction.
The transverse momentum constraint insures that all decays are well
confined within the wire chamber volume.
The angular constraint removes events at large $\cos\theta$ that have
worse resolution and events at small $\cos\theta$ that experience
larger energy loss and multiple scattering.
  These limits were frozen early in the analysis. 
Prior to opening the ``black box'' a study of how the results changed as
each of the fiducial boundaries was moved found the sensitivities to
be very weak.

Figure
\ref{fig:ang_dist} shows the results of the best fit to set B within the
fiducial region for the selected momentum bins.  Set B is one of the
statistically larger sets and is typical of all the sets. Figure~\ref{fig:mc_data}
shows the measured muon decay asymmetry in panels (a) and
(b) while panel (c) presents the difference between the
measured asymmetry and the asymmetry calculated from the best fit MC spectrum
for events within the fiducial region. Panel (d) shows the
difference between the asymmetry within the fiducial as reconstructed and as thrown
for the MC, illustrating that the distortion of the asymmetry by the TWIST detector
is small and essentially momentum independent.

The graphite coated Mylar stopping
target resulted in a time dependence of the muon polarization, $P_{\mu}$,
which prevented the simultaneous determination of a value for $P_{\mu}\xi$
from this data sample. $\langle P_{\mu} \rangle \sim -0.89$ at the time of
decay for the data sets analyzed here.
The graphite coating on the Mylar target was necessary since
the target, also serving as a cathode foil for the two
central MWPC chambers, required a conductive surface. Details regarding the target
can be found in \cite{Henderson}.  Knowing the precise
polarization is not important for extraction of $\delta$, thus the Mylar
target was considered adequate for the current measurement despite the
possibility of depolarizing interactions.

Systematics were studied by employing the fitting technique described above to
fit experimental data samples taken with a systematic parameter set at an
exaggerated level to data taken under ideal conditions. This expresses the
changes in the spectrum shape caused by the systematic effect in terms of
changes in the Michel parameters.  Other systematic sensitivities were
determined by analyzing a data or MC sample with a systematic parameter offset
from its nominal value and fitting to the same sample analyzed with this
parameter at its nominal value.  For the current analysis the largest
uncertainties are for the detector alignment, for the simulation of positron
interactions, and for the chamber response, in particular the time dependent
effects due to gas density changes and to the variability of the cathode foil
positions \cite{Henderson}. The latter parameters were monitored throughout
the data accumulation periods and average values were used in the analysis.
Uncertainties due to the detector alignment were established by analysis of
data and generation of MC with purposely misaligned chambers.  Upper limits
for the positron interaction uncertainties were derived from studies of the
data for muons stopped far upstream and from MC histograms that demonstrated
the distortion of the momentum spectrum due to hard interactions.  Other
important systematic uncertainties for $\delta$ are the stopping target
thickness and the momentum calibration.  The target thickness issue was
studied by varying the thickness of the graphite coating in MC.  
The results of these studies for the
parameter $\delta$ are presented in Tables \ref{tbl:results} and
\ref{tbl:systematics}. The value of $\delta$
for the cloud muon sample is $0.75245 \pm 0.00526$(stat.),
consistent with the results in Table I. The average value of $\rho$ from the
present fits is 0.75044, consistent with the blind analysis result in 
Ref. \cite{Musser}.

The effects of chamber response, momentum calibration and muon beam stability,
which have time dependent components, are treated as data set-dependent
effects with the average(ave) over the four sets used in the $\delta$
evaluation appearing in Table \ref{tbl:systematics}.

\begin{table}
\caption{\label{tbl:results} Results for $\delta$. Each fit has 1887 degrees
of freedom. Statistical and set-dependent systematic uncertainties are
shown. }
\begin{ruledtabular}
\begin{tabular}{lcc}
 Data Set   & $\delta$ & $\chi^2$ \\ \hline
 Set A  & $0.75087\pm0.00156\pm0.00073$ & 1924 \\
 Set B  & $0.74979\pm0.00124\pm0.00055$ & 1880 \\
 1.96 T  & $0.74918\pm0.00124\pm0.00069$ & 1987  \\
 2.04 T  & $0.74908\pm0.00132\pm0.00065$ & 1947 \\
\end{tabular}
\end{ruledtabular}
\end{table}

\begin{table}
\caption{\label{tbl:systematics} Contributions to the systematic uncertainty
for $\delta$.  Average values are denoted by (ave), which are considered
set-dependent when performing the weighted average of data sets.}
\begin{ruledtabular}
\begin{tabular}{lc}
   Effect   & Uncertainty \\ \hline
 Spectrometer alignment & $\pm$0.00061 \\
 Chamber response(ave)  & $\pm$0.00056 \\
 Positron interactions  & $\pm$0.00055 \\
 Stopping target thickness & $\pm$0.00037 \\
 Momentum calibration(ave) & $\pm$0.00029 \\
 Muon beam stability(ave)  & $\pm$0.00010 \\
 Theoretical radiative corrections\cite{Arbuzov} & $\pm$0.00010 \\
 Upstream/Downstream efficiencies & $\pm$0.00004 \\
\end{tabular}
\end{ruledtabular}
\end{table}

We find $\delta$ = 0.74964 $\pm$ 0.00066(stat.) $\pm$ 0.00112(syst.),
consistent with the Standard Model expectation of 3/4.
The central value for $\delta$ was calculated as a weighted
average using a quadratic sum of the statistical and set-dependent uncertainties
for the weights. The final systematic uncertainty is a quadratic sum of
set independent and average values of the set-dependent systematics.
Using this result, our
new value for $\rho$ \cite{Musser}, the previous measurement of
$P_{\mu}\xi\delta/\rho$ \cite{Jodidio}, and the constraint $Q^{\mu}_R\ge0$, it
is possible to establish new 90\% confidence interval limits, $0.9960 <
P_{\mu}\xi \leq \xi < 1.0040$, consistent with the Standard Model value of
1.
This result is more restrictive than the current best measurements
for muons from pion and kaon decays \cite{Beltrami,Imazato}.
In addition, from these same results one finds that
$Q^{\mu}_R < 0.00184$ with 90\% confidence.  This may be combined with
Eq.~(\ref{eq:RHcurrents}) to find new 90\% confidence limits on interactions
that couple right-handed muons to left-handed electrons: $|g^S_{LR}|< 0.086$,
$|g^V_{LR}|< 0.043$, and $|g^T_{LR}|< 0.025$.  The lower limit, $0.9960 <
P_{\mu}\xi$ can be used to determine a new limit on the mass of the possible
right-handed boson, $W_R$, improving the existing lower limit of 406 GeV/c$^2$
(402 GeV/c$^2$ with modern M$_{W_L} = 80.423$ GeV/c$^2$) from \cite{Jodidio}
to 420 GeV/c$^2$ under the assumption of pseudo manifest left-right symmetry.
For nonmanifest left-right symmetric models the limit is $ M_{W_R} g_L/g_R >
380$GeV/c$^2$, where $g_L$ and $g_R$ are the coupling constants
\cite{Herczeg}.  The value of $\delta$ is sensitive to a proposed nonlocal
interaction \cite{Chizhov} that would be represented by a new parameter
$\kappa$. A limit for $\kappa$ may be estimated from our 90\% confidence lower
limit for $\delta$ using the relation $\delta = 3/4(1-6\kappa^2)$.  This
results in $\kappa \leq 0.024$, which compares with $\kappa = 0.013$
\cite{Chizhov} hinted at by $\pi$ decay experiments.

\begin{acknowledgments}
We thank P.A.~Amaudruz, C.A.~Ballard, M.J.~Barnes, S.~Chan, B.~Evans, 
M.~Goyette, K.W.~Hoyle, D.~Maas, J.~Schaapman, J.~Soukup, 
C.~Stevens, G.~Stinson, H.-C.~Walter, and the many undergraduate 
students who contributed to the construction and operation of TWIST.
We also acknowledge many contributions by other professional and technical 
staff members from TRIUMF and collaborating institutions.
Computing resources for the analysis were provided by WestGrid.
This work was supported in part by the Natural Sciences and Engineering 
Research Council and the National Research Council of Canada, the Russian 
Ministry of Science, and the U.S. Department of Energy.
\end{acknowledgments}

\end{document}